\documentclass[]{aa} 
\usepackage{psfig,natbib}
\bibpunct{(}{)}{;}{a}{}{,} 
\begin{document}
\thesaurus{06(08.03.4; 08.05.2; 08.09.2: \object{$\gamma$ Cas};
  08.13.2; 13.09.6)}
\title{The infrared spectrum of the Be star $\gamma$ Cassiopeiae
\thanks{based on observations obtained with ISO, an ESA project with
  instruments funded by ESA Member states (especially the PI
  countries: France, Germany, the Netherlands and the United Kingdom)
  with the participation of ISAS and NASA}}
\author{ S. Hony\inst{1} \and L.B.F.M. Waters\inst{1,2} \and P.A.
  Zaal\inst{1} \and A. de Koter\inst{1} \and J. M. Marlborough\inst{3}
  \and C.E. Millar\inst{3} \and N.R. Trams\inst{4} 
  \and P.W. Morris\inst{1} \and Th. de
  Graauw\inst{5} }
\offprints{S.  Hony (hony@astro.uva.nl)} \institute{ Astronomical
  Institute Anton Pannekoek, University of Amsterdam, Kruislaan 403,
  NL-1098 SJ Amsterdam, The Netherlands 
  \and Instituut voor Sterrenkunde, K.U. Leuven,
  Celestijnenlaan 200B, 3001 B-Heverlee, Belgium 
 \and
  Department of Physics \& Astronomy, University of Western Ontario,
  London, Ontario N6A 3K7, Canada 
  \and Integral Science Operations, Astrophysics Division, Space Science
  Department of ESA, ESTEC, P.O. Box 299, NL-2200 AG Noordwijk, The
  Netherlands
  \and SRON Laboratory for Space
  Reserach, PO Box 800, NL-9700 AV Groningen, The Netherlands 
  } 
\date{Received date; accepted date}
\maketitle
\begin{abstract}
  We present the 2.4$-$45 $\mu$m ISO-SWS spectrum of the Be star
  $\gamma$~Cas (B0.5~IVe). The spectrum is characterised by a thermal
  continuum which can be well fit by a power-law
  S$_{\nu}$~$\propto$~$\nu^{0.99}$ over the entire SWS wavelength
  range.  For an isothermal disc of ionized gas with constant
  opening angle, this correponds to a density gradient
  $\rho$(r)~$\propto$~r$^{-2.8}$.  We report the detection of the
  Humphreys (6-$\infty$) bound-free jump in emission at 3.4 $\mu$m. The
  size of the jump is sensitive to the electron temperature of the gas
  in the disc, and we find T~$\approx$~9\,000~K, i.e. much lower than
  the stellar effective temperature (25\,000-30\,000~K).  The spectrum is
  dominated by numerous emission lines, mostly from \ion{H}{i}, but also some
  \ion{He}{i} lines are detected. Several spectral features cannot be
  identified. The line strengths of the \ion{H}{i} emission lines do not
  follow case B recombination line theory. The line strengths and
  widths suggest that many lines are optically thick and come from an
  inner, high density region with radius 3-5~R$_{*}$ and temperature 
  above that of the bulk of the disc material. 
  Only the
  $\alpha$, $\beta$ and $\gamma$ transitions of the series lines
  contain a contribution from the outer regions. The level populations
  deviate significantly from LTE and are highly influenced by the
  optically thick, local (disc) continuum radiation field.
  The inner disc may be rotating more rapidly than the stellar photosphere. 
\end{abstract}
\section{Introduction}
Be stars are rapidly rotating main sequence or giant stars that are
characterized by the presence of (variable) H$\alpha$ emission, caused
by high-density circumstellar gas. Often the H$\alpha$ line profile is
double-peaked and its width correlates with the projected rotational
velocity (v\,$\sin$\,i) of the underlying star
\citep[e.g.][]{1986A&A...159..276D}. At 
infrared and radio wavelengths, the continuum energy distribution of
Be stars is dominated by free-free and bound-free emission from the
high-density circumstellar gas, which also causes the H$\alpha$
emission.  Direct imaging of the circumstellar material at radio
wavelengths \citep{1992Natur.359..808D} and in the H$\alpha$ line, 
\citep[e.g.][]{1998A&A...332..268S} shows that the gas is
in a disc-like geometry. This flattened geometry is also evident from
the optical continuum linear polarisation
\citep[e.g.][]{1979AJ.....84..812P} due to electron scattering. 

The physical mechanism responsible for the disc-like geometry is
related to the rapid rotation of the star. Models proposed in the
literature include the wind compressed disc model
\citep{1993ApJ...409..429B}, viscocity-driven outflow in the equatorial
regions \citep{1999iau..conf..okazaki}  
and non-radial pulsations \citep{1983ApJ...269..250V}. Which of these
models is correct can be tested by investigating the density and
kinematical structure of the disc. We use the infrared spectral region
to explore the structure of Be star discs. At these wavelengths disc
emission dominates the spectrum, and a rich hydrogen and helium
emission line spectrum is available. The infrared lines probe the
inner regions of the disc, and their strength and width are important
diagnostic tools for the density and kinematical structure of the
disc.  Ground-based infrared spectra of $\gamma$~Cas were previously
reported by e.g. \citet{1983A&A...127..279C}, 
\citet{1985ApJ...290..325L} and \citet{1987ApJ...318..356H}. The
Infrared Space Observatory (ISO) 
\citep{1996A&A...315L..27K} with its Short Wavelength Spectrometer
(SWS) \citep{1996A&A...315L..49D} is very well suited to explore the
infrared part of the spectrum of Be stars. 

In this study we present a preliminary analysis of the ISO-SWS
spectrum of one of the brightest and best studied Be stars in the sky,
$\gamma$~Cas (B0.5IVe). We will show that the \ion{H}{i} emission line
spectrum of $\gamma$~Cas is not well represented by Menzel case B
recombination line theory. Many line fluxes correlate with the local
continuum and are independent of the intrinsic line strength (Einstein
A coefficient). The observed line fluxes and widths suggest that these
lines are formed in an inner region with well-determined size, and
that only the intrinsically strongest lines have a contribution from
outer layers. This paper is organized as follows.  In Sect.~2 we
briefly discuss the observations and data reduction. Section~3 discusses
the continuum and Sect.~4 deals with the line spectrum. Section~5
discusses some implications of our measurements for the structure of
the disc of $\gamma$~Cas.

\section{Observations and data analysis}

\begin{figure*}
  \psfig{figure=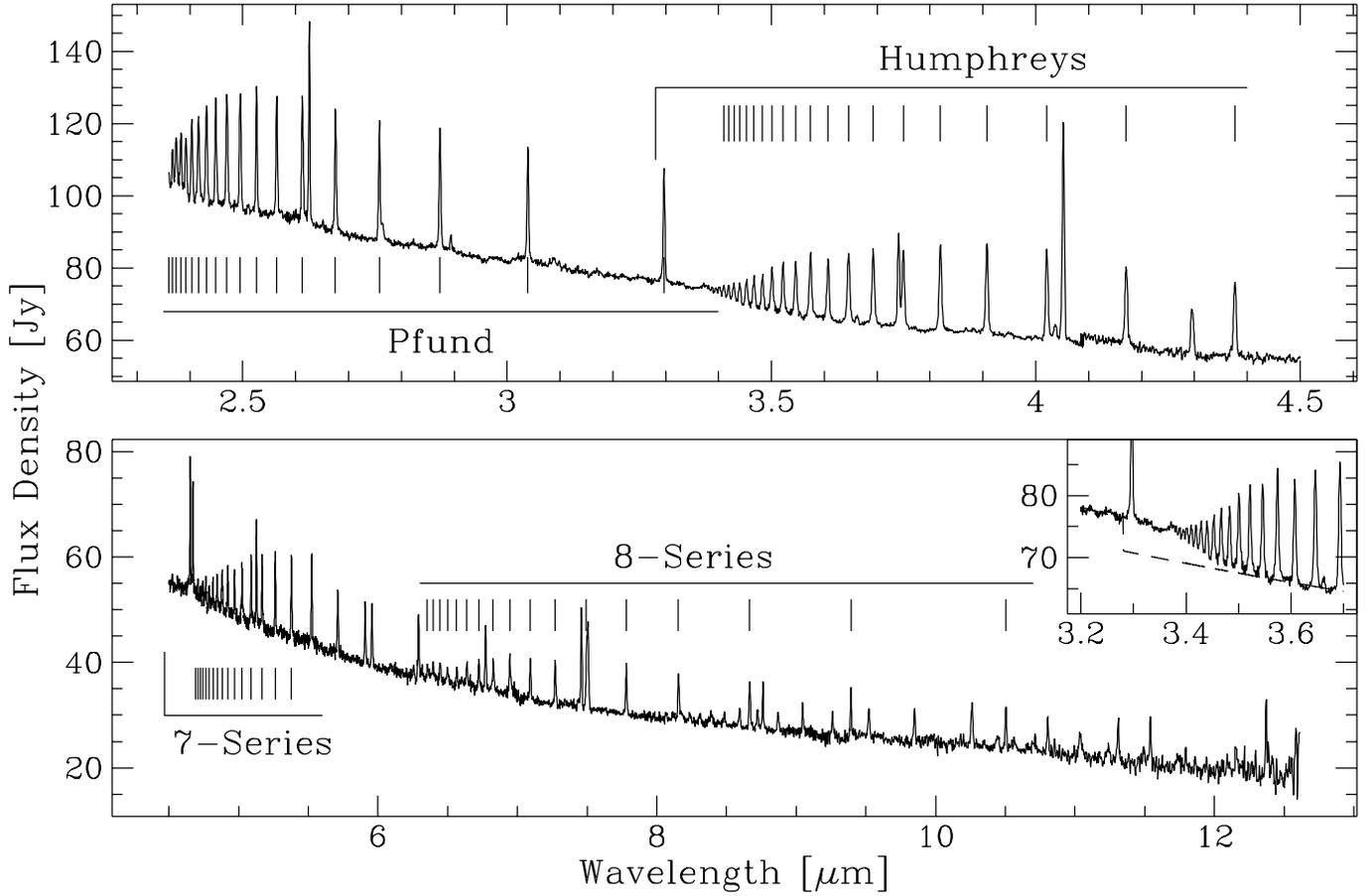,height=12cm}
\caption[spec]{SWS AOT01 speed 4 spectrum of $\gamma$~Cas between 2
  and 12 $\mu$m. The spectrum is dominated by numerous emission lines
  from hydrogen. A few He lines are also observed. The inset shows the
  region aroud the Humphreys jump. The dashed line shows the jump in
  the continuum level due to the jump in bound$-$free opacity near 3.4
  $\mu$m.} 
\label{fig:swsspec}
\end{figure*}

The Be star $\gamma$~Cas was observed with the SWS on board ISO on
July 22nd, 1996, as part of the SWS guaranteed time programme BESTARS.
A full spectral scan (2.4-45 $\mu$m) using Astronomical Observation
Template (AOT) no. 1, speed 4 \citep{1996A&A...315L..49D} was obtained,
while also several AOT02 line scans were taken.  The observations were
reduced using the SWS Interactive Analysis (IA$^3$) software package,
with calibration files equivalent to pipeline version 7.0.  Further processing
consisted of bad data removal and rebinning on an equidistant
wavelength grid. The flux levels are accurate to within 5 per cent for
the wavelengths shortward of 7 $\mu$m. The observations between 7 and
12 $\mu$m (band 2C) suffer from memory effects; this has little
influence on the measured line properties but does increase the
uncertainty of the continuum flux level to 15 per cent. At even longer
wavelengths the signal to noise ratio decreases dramatically and only
the strongest lines can be measured with reasonable accuracy. Most of
the emission lines are partially resolved with a ratio of FWHM to the
FWHM of the instrumental profile between 1.4-3.5. Only 6 of the
emission lines are considered unresolved having this ratio below 1.4. 
Since the SWS instrumental profile is approximately Gaussian
\citep{1996A&A...315L..60V} and the observed lines are well fitted by
Gaussians, we estimate the original line width from:
\begin{equation}
{w_\mathrm{obs}}^2={w_\mathrm{org}}^2+{w_\mathrm{inst}}^2,
\label{eqn:FWHM}
\end{equation}
where $w_\mathrm{obs}$ is the observed FWHM, $w_\mathrm{org}$ is the
original FWHM and $w_\mathrm{inst}$ is the FWHM of the instrumental
profile. The latter value varies with wavelength. To determine
$w_\mathrm{inst}$, we use measured line widths of emission lines
of planetary nebulae; \object{NGC 7027} and \object{NGC 6302},
observed in the same observing mode. No significant line profile
variations are observed. We show the final AOT01 spectrum in
Fig.~\ref{fig:swsspec}. 

\section{The continuum energy distribution}
\begin{figure}[t]
  \centerline{
  \psfig{figure=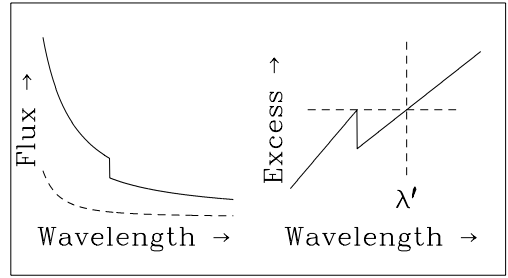,width=8cm}}
  \caption{Schematic representation of the method used to derive the
    temperature from the Humphreys jump. On the left the
    observed flux (drawn line) and the photospheric contribution
    (dotted line) are shown. on the right the normalized excess flux
    is shown. Also indicated is the wavelength($\lambda^{\prime}$)
    longward of the jump where the excess is equal to the excces at
    the jump.}
  \label{fig:humpschema}
\end{figure}

The continuum energy distribution of $\gamma$~Cas at IR wavelengths is
dominated by free-free and bound-free emission from the ionized part
of the circumstellar gas
\citep[e.g.][]{1978ApJ...220..940P,1987A&A...185..206W}. The stellar
contribution to the total 
flux is about 20 per cent at 2.4 $\mu$m, based on extrapolation of a
Kurucz model atmosphere fitted to the UV continuum
\citep{1993A&A...270..355T}. The spectrum can be well represented by a
single power-law, 
S$_{\nu}$~$\propto$~$\nu^{\alpha}$, with
$\alpha$~=~0.99~$\pm$~0.05. This spectral slope is slightly, but
significantly flatter, than that derived by \citet{1987A&A...185..206W},
based on IRAS broad-band photometry taken in 1983. We have used the
simple isothermal disc model of \citet{1986A&A...162..121W} to
estimate the radial density gradient in the disc, assuming a 
power-law $\rho$(r)~=~$\rho_{0}$(r/R$_*$)$^{-n}$, and find
n~=~2.8~$\pm$~0.1.  The value of $\rho_0$ depends on the assumed
opening angle $\theta$ of the disc, as well as on the stellar radius 
and disc temperature. We use R$_*$~=~10~R$_{\sun}$ and
T$_\mathrm{disc}$~=~10$^{4}$~K (see below).  Analysis of the optical
linear polarisation and interferometric imaging of \object{$\zeta$~Tau}
\citep{1997ApJ...477..926W} suggests a half opening angle of
2.5$\degr$. We use a 1$\degr$ half opening angle. The derived density
at the stellar surface is
$\rho_0$~=~3.5$\times$10$^{-11}$~g~cm$^{-3}$; an emission measure 
EM~=~1.5$\times$10$^{61}$ cm$^{-3}$ was found.

There are some wavelength ranges that show a deviation from the
power-law behavior of the continuum discussed above. Near 3.28 $\mu$m
the merging of the emission lines of the Humphreys series, with lower
quantum level $n=6$, causes a jump. This Humphreys jump (seen in
emission), which is similar to the Balmer jump at optical wavelengths,
can be used to derive the average electron temperature of the emitting
region. The difference in flux on both sides of the Humphreys
jump is caused by a discontinuity in bound-free
($\kappa_\mathrm{ff+bf}$) opacity of the gas in the disc. We write for
the total continuum opacity.
\begin{eqnarray}
\kappa_\mathrm{ff+bf} & \propto & \lambda^{2} \times (1-e^{-ch
    /\lambda kT}) / (ch/\lambda kT) \times T^{-3/2} \times \nonumber \\
& & \{g(\lambda ,T) + b(\lambda ,T)\},
\label{eqn:totalkappa}
\end{eqnarray}
where g($\lambda$,T) and b($\lambda$,T) are the free-free and
bound-free gaunt factors, respectively. 
b($\lambda$,T) is a sensitive function of the temperature: the jump in
b($\lambda$,T) (and in flux) increases towards lower electron
temperature. The change in g($\lambda$ ,T) is negligible over the
wavelength range of interest. The jump in b($\lambda$,T) is thus a
diagnostic of the temperature in the disc.
We use the following method to determine the size of this jump:
We define the normalized excess flux as Z$_{\lambda}-$1 =
(F$_{\lambda}$-F$_{\lambda,*}$)/F$_{\lambda,*}$, where F$_{\lambda,*}$
is the stellar photospheric flux, see Fig.~\ref{fig:humpschema}.
Z$_{\lambda}-$1 is normalized to the source function of the gas in the
disc modulo a constant since both the disc and the star radiate in the
Rayleigh-Jeans limit in this wavelength regime and thus have the same
wavelength dependence. 
On the blue side of the discontinuity b($\lambda$,T) has a certain
value, with a corresponding value of $\kappa_\mathrm{ff+bf}$, of 
$\tau_\mathrm{ff+bf}$ for each line of sight through the disc 
and thus a corresponding value of Z$_{\lambda}-$1. Beyond the
discontinuity there is a drop in b($\lambda$,T),
$\kappa_\mathrm{ff+bf}$, $\tau_\mathrm{ff+bf}$ and
Z$_{\lambda}-$1. Since there is a wide range of $\tau_\mathrm{ff+bf}$ for different
lines of sight Z$_{\lambda}-$1 is {\em not} a simple function of
b($\lambda$,T). However $\kappa_\mathrm{ff+bf}$ steadily increases
with wavelength, e.g. Eqn.~\ref{eqn:totalkappa}, and thus there is a
wavelength ($\lambda^{\prime}$) where the loss in b($\lambda$,T) is
compensated by the increase in $\lambda$. At $\lambda^{\prime}$ the
$\kappa_\mathrm{ff+bf}$ is equal to the previous value, so
$\tau_\mathrm{ff+bf}$ and Z$_{\lambda}-$1 are also equal. We can
determine $\lambda^{\prime}$ directly from the observations,
e.g. Fig.~\ref{fig:humpschema}.

Using $\lambda^{\prime}~=~3.470 \pm 0.005~\mu {\mathrm m}$ and the
gaunt factors calculated by \citet{1984A&AS...57..327W} we find an
electron temperature in the 
disc of 9\,500~$\pm$~1\,000~K. This temperature would cause a weak but
measurable jump at the Hansen-Strong series limit (near 4.5 $\mu$m)
but none is observed. However, we note that near this wavelength two
strong \ion{H}{i} lines are present that may mask an otherwise
detectable jump. The disc temperature agrees well with a
density-weighted temperature of 10\,700~K derived by
\citet{1998ApJ...494..715M} from an energy balance 
calculation using the Poeckert \& Marlborough model for the disc of
$\gamma$~Cas. Note that the inner regions of the disc may have
considerably higher temperatures, because this method is insensitive
to contributions of those parts of the highest density parts of the
disc where the continuum is optically thick. (see also Sect.~4).
A second region which deviates from the power-law is near 4.3 $\mu$m.
We cannot find a reasonable identification for this spectral feature.
The CO$_2$ stretch mode band is at 4.27 $\mu$m. However, the
interstellar extinction towards $\gamma$~Cas is very low which 
implies that we must rule out this possibility.

\begin{table*}
\caption{Properties of hydrogenic emission lines. $\lambda$ is the
  peak wavelength of the Gaussian fit. Typical errors on these are
  1/2500th of the wavelength. $width$ is the FWHM of the line after
  deconvolution. $I$ is the integrated line-flux. $cont.$ is the mean
  continuum level underneath the line. The errors on the continuum
  level are dominated by the absolute flux calibration uncertainty of the SWS
  instrument.} 
\begin{tabular}{r @- l r l l l||r @- l r l l l}
\hline
\hline
\multicolumn{2}{c}{(1)}& \multicolumn{1}{c}{(2)}&
\multicolumn{1}{c}{(3)}& \multicolumn{1}{c}{(4)}&
\multicolumn{1}{c}{(5)}& \multicolumn{2}{c}{(6)}&
\multicolumn{1}{c}{(7)}& \multicolumn{1}{c}{(8)}&
\multicolumn{1}{c}{(9)}& \multicolumn{1}{c}{(10)}\\
\multicolumn{2}{c}{$trans.$}& \multicolumn{1}{c}{$\lambda$}&
\multicolumn{1}{c}{$width$}& \multicolumn{1}{c}{$I$}&
\multicolumn{1}{c}{$cont.$}& \multicolumn{2}{c}{$trans.$}&
\multicolumn{1}{c}{$\lambda$}& \multicolumn{1}{c}{$width$}&
\multicolumn{1}{c}{$I$}& \multicolumn{1}{c}{$cont.$}\\

&&\multicolumn{1}{c}{$[\mu m]$}&\multicolumn{1}{c}{$[kms^{-1}]$}& 
\multicolumn{1}{c}{$[W/m^2]$}& \multicolumn{1}{c}{$[Jy]$}& 
&&\multicolumn{1}{c}{$[\mu m]$}& \multicolumn{1}{c}{$[kms^{-1}]$}&
\multicolumn{1}{c}{$[W/m^2]$}& \multicolumn{1}{c}{$[Jy]$}\\
\hline
 5&4  &  4.052 & 254 $\pm$ 42 & 5.45e-14 $\pm$ 6\%&  61$\pm$ 6\%& 10&7  &  8.761 & 249 $\pm$ 39 & 3.70e-15 $\pm$10 \% & 24$\pm$10\%\\
 6&4  &  2.626 & 226 $\pm$ 56 & 7.14e-14 $\pm$ 5  &  91$\pm$ 5  & 12&7  &  6.772 & 364 $\pm$ 40 & 8.21e-15 $\pm$ 8    & 35$\pm$ 8  \\
 6&5  &  7.460 & 220 $\pm$ 65 & 9.51e-15 $\pm$ 8  &  28$\pm$ 8  & 13&7  &  6.292 & 414 $\pm$ 41 & 1.02e-14 $\pm$ 7    & 37$\pm$ 7  \\
 7&5  &  4.654 & 200 $\pm$ 43 & 1.58e-14 $\pm$ 6  &  56$\pm$ 6  & 14&7  &  5.957 & 303 $\pm$ 64 & 9.11e-15 $\pm$ 7    & 40$\pm$ 7  \\
 8&5  &  3.741 & 267 $\pm$ 48 & 2.69e-14 $\pm$ 6  &  64$\pm$ 5  & 15&7  &  5.712 & 376 $\pm$ 56 & 1.12e-14 $\pm$ 7    & 42$\pm$ 7  \\
 9&5  &  3.297 & 284 $\pm$ 23 & 3.33e-14 $\pm$ 5  &  77$\pm$ 5  & 16&7  &  5.525 & 393 $\pm$ 13 & 1.34e-14 $\pm$ 7    & 45$\pm$ 6  \\
10&5  &  3.039 & 273 $\pm$ 32 & 3.73e-14 $\pm$ 5  &  80$\pm$ 5  & 17&7  &  5.380 & 367 $\pm$ 16 & 1.21e-14 $\pm$ 7    & 46$\pm$ 6  \\
11&5  &  2.873 & 309 $\pm$ 33 & 4.53e-14 $\pm$ 5  &  84$\pm$ 5  & 18&7  &  5.264 & 335 $\pm$ 18 & 1.13e-14 $\pm$ 7    & 47$\pm$ 6  \\
12&5  &  2.758 & 313 $\pm$ 36 & 4.73e-14 $\pm$ 5  &  86$\pm$ 5  & 19&7  &  5.169 & 390 $\pm$ 16 & 1.09e-14 $\pm$ 7    & 48$\pm$ 6  \\
13&5  &  2.675 & 324 $\pm$ 37 & 5.37e-14 $\pm$ 5  &  88$\pm$ 5  & 20&7  &  5.091 & 396 $\pm$ 17 & 1.16e-14 $\pm$ 6    & 49$\pm$ 6  \\
14&5  &  2.613 & 347 $\pm$ 12 & 5.20e-14 $\pm$ 5  &  95$\pm$ 5  & 21&7  &  5.026 & 457 $\pm$ 15 & 1.11e-14 $\pm$ 6    & 49$\pm$ 6  \\
15&5  &  2.564 & 340 $\pm$ 13 & 5.38e-14 $\pm$ 5  &  95$\pm$ 5  & 22&7  &  4.971 & 441 $\pm$ 16 & 9.04e-15 $\pm$ 7    & 50$\pm$ 6  \\
16&5  &  2.526 & 344 $\pm$ 14 & 5.68e-14 $\pm$ 5  &  96$\pm$ 5  & 23&7  &  4.923 & 425 $\pm$ 17 & 8.34e-15 $\pm$ 7    & 50$\pm$ 6  \\
17&5  &  2.495 & 379 $\pm$ 13 & 5.53e-14 $\pm$ 5  &  97$\pm$ 5  & 24&7  &  4.883 & 551 $\pm$ 13 & 8.86e-15 $\pm$ 7    & 51$\pm$ 6  \\
18&5  &  2.470 & 407 $\pm$ 12 & 5.66e-14 $\pm$ 5  &  98$\pm$ 5  & 25&7  &  4.847 & 441 $\pm$ 17 & 6.69e-15 $\pm$ 7    & 51$\pm$ 6  \\
19&5  &  2.449 & 386 $\pm$ 13 & 5.17e-14 $\pm$ 5  &  98$\pm$ 5  & 26&7  &  4.817 & 420 $\pm$ 18 & 5.18e-15 $\pm$ 8    & 52$\pm$ 6  \\
20&5  &  2.431 & 390 $\pm$ 13 & 4.83e-14 $\pm$ 5  &  99$\pm$ 5  & 27&7  &  4.789 & 455 $\pm$ 17 & 4.39e-15 $\pm$ 8    & 52$\pm$ 6  \\
21&5  &  2.416 & 447 $\pm$ 12 & 4.79e-14 $\pm$ 5  &  99$\pm$ 5  & 28&7  &  4.765 & 551 $\pm$ 14 & 5.45e-15 $\pm$ 8    & 52$\pm$ 6  \\
22&5  &  2.404 & 417 $\pm$ 13 & 4.24e-14 $\pm$ 5  & 100$\pm$ 5  & 29&7  &  4.744 & 706 $\pm$ 11 & 5.59e-15 $\pm$ 8    & 52$\pm$ 6  \\
23&5  &  2.393 & 414 $\pm$ 13 & 3.20e-14 $\pm$ 5  & 100$\pm$ 5  & 12&8  & 10.504 & 282 $\pm$ 22 & 2.88e-15 $\pm$13    & 20$\pm$12  \\
24&5  &  2.383 & 491 $\pm$ 11 & 4.03e-14 $\pm$ 5  & 101$\pm$ 5  & 13&8  &  9.393 & 204 $\pm$ 41 & 2.97e-15 $\pm$11    & 23$\pm$11  \\
 7&6  & 12.371 & 313 $\pm$ 13 & 4.34e-15 $\pm$15  &  15$\pm$15  & 14&8  &  8.666 & 339 $\pm$ 29 & 4.40e-15 $\pm$10    & 24$\pm$10  \\
 9&6  &  5.908 & 295 $\pm$ 67 & 8.81e-15 $\pm$ 7  &  41$\pm$ 7  & 15&8  &  8.156 & 345 $\pm$ 33 & 4.69e-15 $\pm$10    & 26$\pm$ 9  \\
10&6  &  5.129 & 325 $\pm$ 20 & 1.46e-14 $\pm$ 6  &  48$\pm$ 6  & 16&8  &  7.781 & 291 $\pm$ 44 & 4.87e-15 $\pm$ 9    & 27$\pm$ 9  \\
11&6  &  4.673 & 304 $\pm$ 27 & 1.57e-14 $\pm$ 6  &  56$\pm$ 6  & 18&8  &  7.272 & 252 $\pm$ 60 & 4.24e-15 $\pm$ 9    & 29$\pm$ 8  \\
12&6  &  4.377 & 326 $\pm$ 30 & 1.98e-14 $\pm$ 6  &  57$\pm$ 6  & 19&8  &  7.093 & 321 $\pm$ 49 & 4.83e-15 $\pm$ 9    & 30$\pm$ 8  \\
13&6  &  4.171 & 348 $\pm$ 32 & 2.33e-14 $\pm$ 6  &  61$\pm$ 6  & 20&8  &  6.947 & 484 $\pm$ 28 & 5.67e-15 $\pm$ 8    & 34$\pm$ 8  \\
14&6  &  4.021 & 304 $\pm$ 35 & 2.52e-14 $\pm$ 6  &  60$\pm$ 6  & 21&8  &  6.827 & 445 $\pm$ 32 & 4.48e-15 $\pm$ 9    & 35$\pm$ 8  \\
15&6  &  3.908 & 332 $\pm$ 35 & 2.73e-14 $\pm$ 6  &  62$\pm$ 5  & 22&8  &  6.724 & 376 $\pm$ 39 & 3.67e-15 $\pm$ 9    & 36$\pm$ 8  \\
16&6  &  3.820 & 350 $\pm$ 35 & 2.77e-14 $\pm$ 6  &  63$\pm$ 5  & 23&8  &  6.639 & 421 $\pm$ 36 & 3.47e-15 $\pm$ 9    & 36$\pm$ 7  \\
17&6  &  3.750 & 360 $\pm$ 35 & 2.68e-14 $\pm$ 6  &  64$\pm$ 5  & 24&8  &  6.566 & 528 $\pm$ 29 & 3.82e-15 $\pm$ 9    & 35$\pm$ 7  \\
18&6  &  3.693 & 372 $\pm$ 35 & 2.63e-14 $\pm$ 6  &  65$\pm$ 5  & 26&8  &  6.447 & 671 $\pm$ 24 & 4.53e-15 $\pm$ 8    & 36$\pm$ 7  \\
19&6  &  3.646 & 392 $\pm$ 34 & 2.54e-14 $\pm$ 6  &  65$\pm$ 5  & 15&9  & 11.539 & 275 $\pm$ 18 & 2.71e-15 $\pm$14    & 17$\pm$14  \\
20&6  &  3.607 & 344 $\pm$ 40 & 2.08e-14 $\pm$ 6  &  66$\pm$ 5  & 16&9  & 10.805 & 301 $\pm$ 19 & 2.26e-15 $\pm$14    & 20$\pm$13  \\
21&6  &  3.574 & 387 $\pm$ 36 & 2.06e-14 $\pm$ 6  &  67$\pm$ 5  & 17&9  & 10.261 & 431 $\pm$ 15 & 4.07e-15 $\pm$12    & 21$\pm$12  \\
22&6  &  3.546 & 380 $\pm$ 38 & 1.95e-14 $\pm$ 6  &  67$\pm$ 5  & 18&9  &  9.848 & 292 $\pm$ 25 & 2.36e-15 $\pm$12    & 22$\pm$11  \\
23&6  &  3.522 & 408 $\pm$ 36 & 1.69e-14 $\pm$ 6  &  67$\pm$ 5  & 19&9  &  9.522 & 337 $\pm$ 24 & 2.21e-15 $\pm$12    & 23$\pm$11  \\
24&6  &  3.501 & 466 $\pm$ 32 & 1.75e-14 $\pm$ 6  &  68$\pm$ 5  & 20&9  &  9.261 & 364 $\pm$ 23 & 2.18e-15 $\pm$12    & 22$\pm$10  \\
25&6  &  3.483 & 476 $\pm$ 31 & 1.49e-14 $\pm$ 6  &  68$\pm$ 5  & 21&9  &  9.047 & 283 $\pm$ 32 & 2.05e-15 $\pm$12    & 23$\pm$10  \\
26&6  &  3.467 & 438 $\pm$ 34 & 1.18e-14 $\pm$ 6  &  68$\pm$ 5  & 22&9  &  8.870 & 422 $\pm$ 22 & 1.96e-15 $\pm$12    & 24$\pm$10  \\
27&6  &  3.452 & 621 $\pm$ 24 & 1.40e-14 $\pm$ 6  &  69$\pm$ 5  & 23&9  &  8.721 & 416 $\pm$ 24 & 1.88e-15 $\pm$12    & 24$\pm$10  \\
28&6  &  3.440 & 639 $\pm$ 24 & 1.29e-14 $\pm$ 6  &  69$\pm$ 5  & 24&9  &  8.594 & 375 $\pm$ 27 & 1.82e-15 $\pm$12    & 25$\pm$10  \\
29&6  &  3.429 & 641 $\pm$ 24 & 1.08e-14 $\pm$ 6  &  69$\pm$ 5  & 25&9  &  8.485 & 430 $\pm$ 24 & 1.36e-15 $\pm$14    & 25$\pm$ 9  \\
 9&7  & 11.309 & 394 $\pm$ 13 & 2.96e-15 $\pm$14  &  17$\pm$13  & 20&10 & 12.156 & 439 $\pm$  9 & 1.91e-15 $\pm$16    & 16$\pm$15  \\
\hline
\hline
\end{tabular}
\label{tab:hlines}
\end{table*}

\section{The emission lines}

The entire SWS spectral region, but especially bands 1 and 2 (2.4-12
$\mu$m), is dominated by strong and partially resolved emission lines.
The vast majority of these lines are \ion{H}{i} recombination lines. We find
the series limit of the Humphreys and Hansen-Strong series, as well as
those from lower levels 8, 9 and 10. Several lines with lower levels
above 10 were also identified. We find a few \ion{He}{i} lines (notably
at 2.4861, 2.5729, 4.2960 and 4.0367 $\mu$m). No forbidden
lines could be found, although several emission features are
unidentified and could perhaps be attributed to forbidden lines. The
strongest unidentified line is at 2.8934 $\mu$m, close to the 2.8964
[\ion{Ni}{ii}] line.  The lack of fine-structure lines is consistent with the
optical spectrum of $\gamma$~Cas.  It is likely that the high density
of the ionized gas causes collisional de-excitation of the
fine-structure transitions.  This situation is markedly different for
the hypergiant \object{P~Cygni}, whose infrared spectrum is also dominated by
\ion{H}{i} recombination lines from circumstellar gas, but which also shows
prominent emission from e.g.  [\ion{Fe}{ii}], [\ion{Ni}{ii}] and
[\ion{Si}{ii}] in its ISO-SWS spectrum \citep{1996A&A...315L.229L}.

\begin{figure}[t]
  \psfig{figure=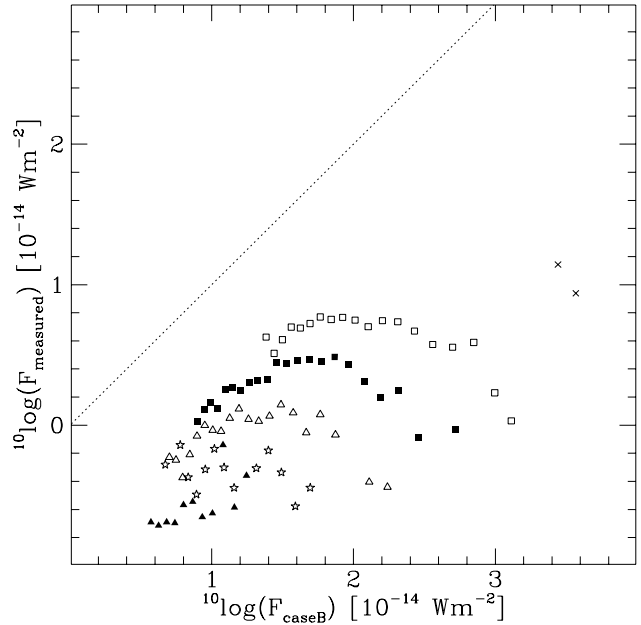,width=8.8cm}
  \caption[spec]{Comparison of measured line strengths with predictions
    based on optically thin case B recombination theory.  The dotted line
    denotes the locus where the measured fluxes would equal the
    predictions.
    The crosses correspond to Brackett $\alpha$ and $\beta$,
    the open squares to the Pfund series,
    the filled squares to the Humphreys series,
    the open triangles to the Hansen-Strong series, 
    the stars to $\mathrm{n}=8$, 
    and the filled triangles to $\mathrm{n}=9$.}
  \label{fig:caseB}
\end{figure}

\begin{figure}[t]
\psfig{figure=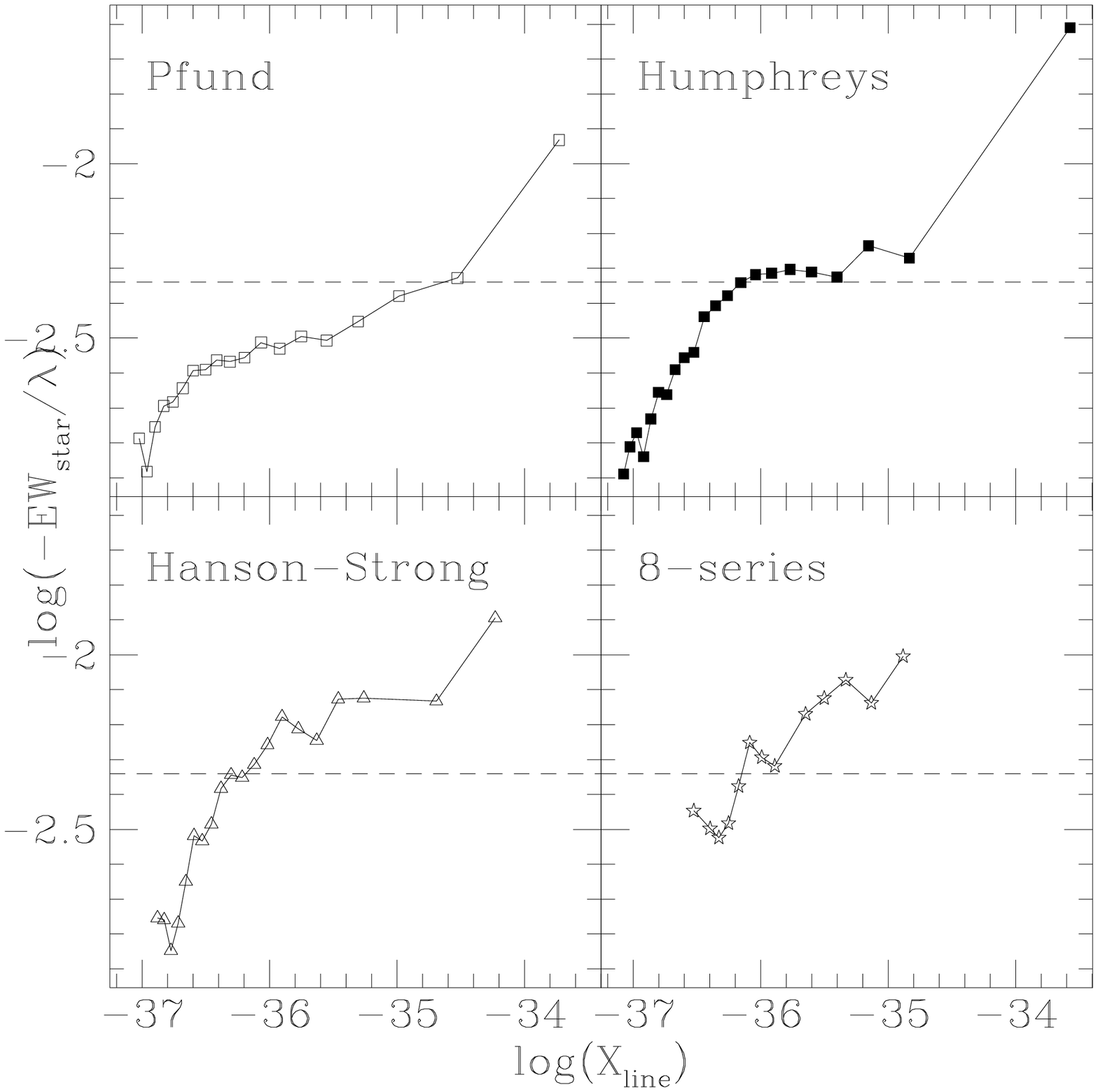,width=8.8cm}
\caption[cog]{Measured line fluxes, expressed in equivalent 
  width (EW/$\lambda$) with respect to the stellar continuum
  versus line opacity. The line fluxes are
  independent of line strength between log X$_\mathrm{line}$~$\approx$~$-$36 
and $-$35. 
}
\label{fig:coglines}
\end{figure}

We have measured the line fluxes with respect to both the {\em local}
continuum and the {\em stellar} photospheric
continuum at line centre. The measured line properties are given in
Table~\ref{tab:hlines}. In Fig.~\ref{fig:coglines} we show the
resulting curve of growth for the emission lines, where we plot 
the equivalent width EW divided by
wavelength versus log X$_\mathrm{line}$. This last quantity is 
proportional to the line optical depth \citep{1995A&A...299..574Z}.
The observed line fluxes (Fig.~\ref{fig:caseB}) are all much smaller 
(a factor 10 or more) than expected on the basis of the continuum
emission measure determination and the optically thin case B
recombination line predictions \citep{1987MNRAS.224..801H}. This shows
that even for the weakest lines the emission is optically thick: the
combined line and continuum opacity is much larger than 1. This is not
surprising given the shape of the continuum energy distribution;
\citet{1991A&A...244..120W} show that the continuum
becomes optically thick near 0.8 $\mu$m. 

The shape of the empirical curve of growth is very different from that
expected on the basis of LTE line formation in a rotating, partially
optically thick disc. Model calculations show that under these
conditions the curve of growth has a linear part (where line emission
is optically thin and proportional to line strength), and a power-law
part whose slope depends on the radial density gradient of the gas
\citep{1995A&A...299..574Z}, see also \citet{1952ZA.....30...96W}. The
observed curve of growth however shows a very 
steep rise of EW/$\lambda$ with X$_\mathrm{line}$ up to
log~X$_\mathrm{line}$~$\approx$~$-$36.5, followed by a flat part for 
$-$36.5~$<$~log~X$_\mathrm{line}$~$<-$35.0 whose slope is close to zero,
i.e.  much smaller than 0.4 expected on the basis of the density
gradient derived from the continuum free-free and bound-free excess.
Finally, a rising part for the strongest lines in the spectrum
(log~X$_\mathrm{line}$~$>-$35.0) is seen. The flat part of the curve of
growth is surprising and suggests that these lines have saturated and
are formed in a region with a well-defined outer radius. We note that
the size of this region is not the same for every series, but
increases with lower quantum level.
 
\begin{figure*}
\psfig{figure=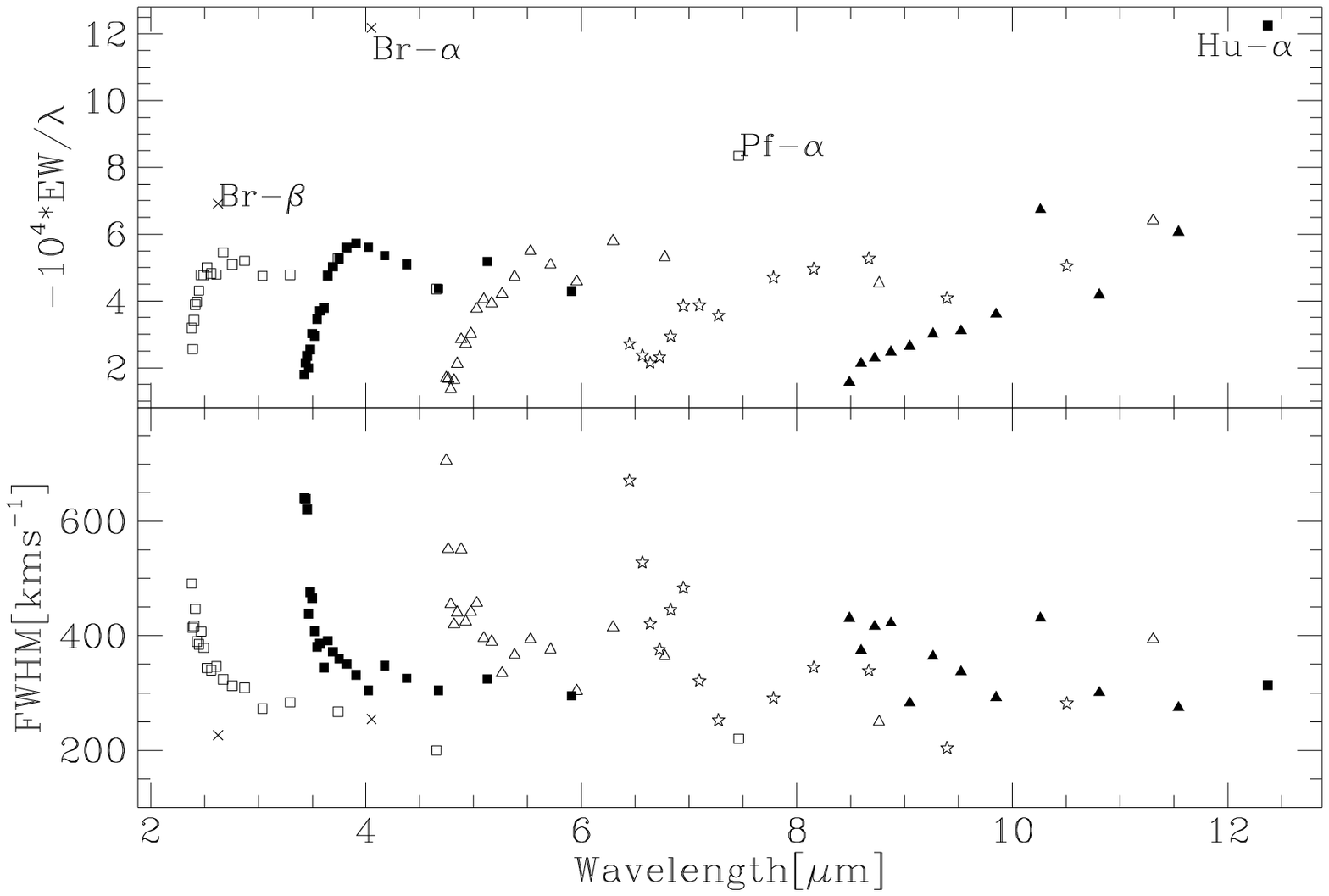,height=12cm}
\caption[cog]{{\bf Top panel}: EW/$\lambda$ of the \ion{H}{i} lines in
  the ISO-SWS spectrum of $\gamma$~Cas as a function of
  wavelength. Symbols are the same as in Fig.~\ref{fig:caseB}. The EWs
  have been determined with respect to the local continuum. The EWs
  for all series behave in a similar way. For moderately strong lines
  the EW/$\lambda$ does not depend on wavelength and is about
  5$\times$10$^{-3}$ irrespective of the series. Only the strongest
  lines of each series deviate from this trend. {\bf Bottom panel}:
  observed full width at half maximum of the \ion{H}{i} lines versus
  wavelength. The symbols refer to the same series as in the upper
  panel. The lines become narrower with increasing line strength as
  expected in a rapidly rotating disc in which the rotational velocity
  decreases with distance. The FWHM does not decrease further for
  moderately strong lines. These lines also show a constant
  EW/$\lambda$ (upper panel).}
\label{fig:vel}
\end{figure*}

In order to understand better the nature of the line formation in
$\gamma$~Cas, we show in Fig.~\ref{fig:vel} the line strength
$\overline{EW}$/$\lambda$, where $\overline{EW}$ is the line
equivalent width measured with respect to the {\em local} continuum,
as a function of wavelength. Also shown in
Fig.~\ref{fig:vel} is the line FWHM versus wavelength. Both
$\overline{EW}$/$\lambda$ and the FWHM show a characteristic
wavelength dependence: for each series, $\overline{EW}$/$\lambda$
increases and then saturates to a value of 5-6$\times$10$^{-4}$, while the
line FWHM decreases and reaches a roughly constant value of 250-300
km~s$^{-1}$.  In other words, the lines reach a constant line over
continuum ratio of about 1.3 at a constant FWHM of 250-300
km~s$^{-1}$.  Only the $\alpha$ and $\beta$ lines of each series
deviate from this behaviour: their line over continuum ratio is larger
and their FWHM smaller.

The decrease of FWHM with increasing line strength is expected for
lines formed in a rotating disc in which the rotational velocity
decreases with distance; as more line flux is coming from the outer,
more slowly rotating regions, the line width will decrease. This is
direct proof of the rotating nature of the line emitting region. It is
remarkable that the weakest lines of each series have a very high FWHM
of more than 550 km~s$^{-1}$. Such high velocities are not expected
given the photospheric v\,$\sin$\,i of 230 km~s$^{-1}$
\citep{1982ApJS...50...55S}. 
This suggests that the inner disc is rotating more rapidly than the
star. However, line broadening due to electron scattering could also
cause these large line widths, but we do not observe prominent
electron scattering wings.

If we assume that the line and continuum source functions are equal, no
line emission from the layers with $\tau_\mathrm{ff}$~$>$~1 should be
detectable. However, the large line width of the weakest lines
strongly suggests that this line emission is originating from rapidly
rotating parts of the disc. The fact that the line FWHM decreases with
increasing line strength shows that the broad, weak lines are formed
closest to the star.  In the 2.5 to 7 $\mu$m wavelength region, where
most of the weakest lines in Fig.~\ref{fig:vel} are located, the
continuum is optically thick out to a radius of 2.5 to 3 R$_{*}$. Assuming
Keplerian rotation in the disc, the part of the disc which is not
optically thick for continuum radiation rotates at projected speeds less than
about v$_{0}$\,$\sin$\,i/1.7, where v$_0$ is the Keplerian velocity of
the disc at the stellar surface. Using reasonable values for the mass
and radius for a B0.5~IV star of R$_{*}$=10\,R$_\odot$ and
M$_{*}$=15\,M$_\odot$ , we find a maximum speed of about 220
km~s$^{-1}$. Emission lines in the disc whose
source function is equal to that of the continuum therefore should
have \emph{FWZI} less than 440 km~s$^{-1}$, but even the FWHM
of the weak lines is significantly higher than this. It is unlikely
that rotational velocities are as high 
as 250-300 km~s$^{-1}$ as far out as 2.5-3~R$_{*}$, unless the 
rotational velocity field deviates significantly from Keplerian.
We conclude that we detect line emission from those parts of the disc
that are optically thick for continuum radiation, and hence the line
source function in these inner regions must exceed that of the continuum
(which is the Planck function at the local electron temperature).
Several effects can cause such a larger source function: NLTE level
populations, or an enhanced electron temperature near the upper part
of the disc. If the latter effect is important, temperatures in the
line forming regions may be as much as 30 per cent higher than in the
bulk of the disc, since the stronger lines are 1.3 times the
continuum.

As pointed out above, the observed line fluxes of every series first
increase with intrinsic line strength, but then reach a constant value
with respect to the local continuum, both in $\overline{EW}$/$\lambda$
and in the line to continuum ratio. This effect occurs for
transitions from upper levels roughly between 15 and about 10, to
lower levels between 9 and 5. The line fluxes are no longer determined
by the intrinsic line strength, but by {\em the local continuum!} The
contribution from outer regions, with lower rotational velocities, is
absent. This can be seen from the FWHM's of the lines with constant line
over continuum ratio, that no longer decrease but are constant at about
250-300 km~s$^{-1}$. This could be due to a finite outer radius of the
disc. However, the ISO-SWS spectrum as well as ISO-PHOT photometry at
90 $\mu$m (Trams et al., in preparation) show that there is no change
in continuum slope over a wide wavelength range. Such a change in
slope would be expected in the case of a finite outer radius. We
conclude that the lack of line emission from the outer regions cannot
be due to a finite outer disc radius. We note that the fact that the
line to continuum ratio is constant for certain lines, irrespective
of the strength of the local continuum and thus of the size of the
continuum emitting region, also argues against a finite outer radius.

The lack of line flux from the outer regions therefore must be due to
a change (decrease) in the line source function compared to the inner
regions.  For a certain range of intermediate energy levels in the
atom, the optically thick IR free-free disc radiation field in the
inner regions is able to maintain significant level populations, but
these levels rapidly become de-populated when the local free-free
continuum becomes optically thin.  Therefore the line flux does not
increase with intrinsic line strength (as do the weaker lines from
high upper levels), but reaches a line over continuum ratio determined
by the ratio of source functions of line and continuum.

\section{Discussion}

\citet{1987ApJ...318..356H} discuss the near-IR
spectrum of $\gamma$~Cas and 
note that the line flux ratios of the weak Pfund lines to the Br$\gamma$
line are very difficult to reconcile with those of Br$\gamma$ and
Br$\alpha$ for any reasonable choice of radial density gradient in the disc.
We confirm this result and show that it is a general
property of the line fluxes of $\gamma$~Cas to behave in a highly
non-standard way, probably caused by a combination of non-LTE and
temperature effects. We do not confirm the lack of correlation between
line strength and line width noted by \citet{1987ApJ...318..356H};  
the ISO data clearly show such a correlation for all series in our 
spectrum. This may be due to the Earth atmosphere affecting the estimate
of the weak high level Pfund series lines in the spectrum of Hamann \& Simon.
The line widths of their stronger lines (e.g. Br$\alpha$) agree well 
with those of our spectrum. They also agree well with the line widths
derived from high-resolution spectra published by
\citet{1985ApJ...290..325L} and \citet{1983A&A...127..279C}.  

The FWHM of the weakest \ion{H}{i} lines (more than 500 km~s$^{-1}$) 
significantly exceeds that of the photospheric 
v\,$\sin$\,i of 230 km~s$^{-1}$. While
electron scattering may affect the width of the strongest lines
(especially at short wavelengths where free-free opacities are small
compared to the electron scattering opacity) the IR lines do not show
evidence for prominent electron scattering wings. This suggests that the
line width is mainly due to kinematical broadening, which would imply
that the disc is rotating more rapidly than the underlying star 
(ignoring errors in v\,$\sin$\,i). This can only occur if some mechanism adds
angular momentum to the material injected into the disc. We note that
\citet{1987A&A...173..299H} found line widths of weak
\ion{Fe}{ii} emission lines in the optical spectra of Be stars that also
significantly exceeded the width expected on the basis of the observed
v\,$\sin$\,i. Transfer of angular momentum may lead to spin-down of the star
\citep{1998A&A...333L..83P}. Using the formalism given by
\citet{1998A&A...333L..83P}, and assuming a full opening angle of 2
degrees and $\rho_0$ of 3$\times$10$^{-11}$ 
g~cm$^{-3}$, the outflow velocity in the disc near the star can be of
the order of 1 km~s$^{-1}$ without significant spin-down of the star
during its main sequence life time. Such a value is in agreement with
the observed line shape. 

The picture which emerges from our analysis of the infrared spectrum 
of $\gamma$~Cas is that of a circumstellar region of very high density, 
perhaps exceeding 3~10$^{-11}$ g~cm$^{-3}$, which is rotating rapidly and
whose rotational velocity decreases with distance. The rotational
velocity in the disc near the star exceeds that of the photosphere.
This region is heated
by radiation from the central star, and the surface layers which
directly absorb the stellar radiation field have higher temperatures
than regions closer to the equatorial plane of the disc.
While the weak lines originate from these dense, warm regions,
only the strongest, $\alpha$ and $\beta$, lines of each series 
probe the outer portions of the disc.

\begin{acknowledgements}LBFMW and AdK acknowledge
financial support from an NWO {\em Pionier} grant. JMM acknowledges
support from an NSERC grant. JMM and LBFMW acknowledge
financial support from a NATO Collaborative Resarch Grant (CRG.941220). 
This work was supported by NWO Spinoza grant 08-0 to 
E.P.J. van den Heuvel. CEM acknowledges financial support from an
NSERC postgraduate scholarship.
\end{acknowledgements}

\bibliographystyle{NBaa}
\bibliography{articles}

\end{document}